\documentclass[12pt]{article}
\usepackage{graphicx}
\graphicspath{ {./images/} }

\usepackage{amssymb,amsmath}
\usepackage{cite}

\begin{document}

\begin{center}
\phantom{0}
\vskip -20mm

\vskip 10mm

{\bf
Alexander Andrianov,
\footnote{St.Petersburg state university, St.Petersburg, Russia, email: sashaandrianov@gmail.com}
Danila Lyozin,
\footnote{St.Petersburg state university, St.Petersburg, Russia, email:  danilalyozin@yandex.ru}
and Artem Starodubtsev
\footnote{St.Petersburg state university, St.Petersburg, Russia,
Leonard Euler Saint-Petersburg international mathematical institute, St.Petersburg, Russia, email:  artemstarodubtsev@gmail.com}}

\vskip 0.5em

{\bf
The spectrum of states of Ba\~{n}ados-Teitelboim-Zanelli black hole formed by a collapsing dust shell
}

\end{center}

\begin{abstract}
We perform canonical analysis of an action in which 2+1-dimensional gravity with negative cosmological constant is coupled to
cylindrically symmetric dust shell. The resulting phase space is finite dimensional having geometry of $SO(2,2)$ group manifold.
Representing the Poisson brackets by commutators results in the algebra of observables which is a quantum double $D(SL(2)_q)$.
Deformation parameter $q$ is real when the total energy of the system is below the threshold of a black hole formation, and a root of
unity when it is above. Inside the black hole the spectra of the shell radius and time operator are discrete and take on a
finite set of values. The Hilbert space of the black hole is thus finite-dimensional.
\end{abstract}


\section{Introduction}
One of the key questions about  black hole evolution is whether the process of its formation and subsequent evaporation is unitary or not.
Classical results predict loss of unitarity, on the other hand in \cite{thooft} it was shown that with certain assumptions about
the wavefunction unitarity could be restored. The analysis of the above work was done at the semiclassical level, where quantum fields
evolve at a classical spacetime background.

This is not the only way to proceed in the absence of the full theory of gravity. One can consider symmetry reduced models, which contain only
finite number of degrees of freedom, but can nevertheless describe the full process of black hole formation and subsequent evaporation.
For such models full quantum theory can be constructed. This was done for spherically symmetric dust shell, coupled to gravity \cite{vberezin}.
However, because of the complexity of the model, some approximations had to be used. This model becomes simpler in 2+1 spacetime dimensions \cite{aes1}.

In our previous paper \cite{aes2}, the dynamics of a dust shell near the Ba\~{n}ados-Teitelboim-Zanelli  black hole \cite{btz} horizon was considered.
In that region, one can use approximation of small translations, and the momenta non-commutativity could be neglected.
This allowed us to study dynamics using representations of classical groups. Transition amplitudes between different regions
of the Penrose diagram near horizon were found. Nevertheless the conclusion was made that this information is not sufficient
to describe the evolution of a black hole in full. Classically, for a large black hole, a collapsing shell
continue to move towards the central singularity. To   get the full picture, one needs to know what happens in the depth of a black
hole where quasi-classical approximation is not valid.

If we want to explore the interior of the black hole in depth, in particular to study what happens
near singularity, the approximation  used in \cite{aes2} is no longer valid.
The momentum space becomes non-commutative, and in that regime we have the full-fledged  quantum group structure.

In section \ref{sec2}, we sketch the derivation of Poisson-Lie structure from gravity action. The momentum space is given by $SL(2)$ group manifold,
coordinate space by translations in $ADS^3$, and Poisson brackets by canonical Drinfeld $r$-matrix.

In section  \ref{sec3}, we describe the result of quantization of Poisson-Lie structure of the previous section - quantum double $D(SL(2)_q)$ and its real forms.
It turns out that both real forms $D(SU(1,1)_q)$, $q^*=q$ and $D(SL(2)_q)$, $q^*=q^{-1}$ are realized in this model, depending on the overall energy of the shell.

In section  \ref{sec4}, we find the spectra of the time operator and the shell radius inside the black hole. The spectra of both operators turn out to be both
discrete and bounded, and the Hilbert space of a black hole is thus finite-dimensional.

Few open problems like calculation of transition amplitudes between the states found and application of this model to black hole entropy explanation are discussed.

%%%%%%%%%%%%%%%%%%%%%%%%%%%%%%%%%%%%%%%%%%
\section{Symplectic form and classical r-matrix}\label{sec2}

We start with describing 2+1-dimensional gravity action with negative cosmological constant, $\Lambda<0$, coupled to circular dust shell.
The basic variable is SO(2,2)-connection $A^{AB}_\mu$, where $A,B=0..3$. This connection contains both:  the triad $A^{3a}_\mu=e^a_{\mu}/l$, where $l=1/\sqrt{|\Lambda|}$, and  the Lorentzian connection $A^{ab}_\mu=\omega^{ab}_\mu$, where $a,b=0..2$.
The total action consists of gravity action in the Chern-Simons form and the shell action
\begin{equation}\label{action}
S= \frac{l}{8\pi} \int_M d^3 x \epsilon^{\mu \nu \rho}\langle A_\mu, (\partial_\nu A_\rho +\frac{2}{3} A_\nu A_\rho)  \rangle + S_{shell},
\end{equation}
where $A_\mu=\Gamma_{AB}A^{AB}_\mu$ is $so(2,2)$ connection, $\Gamma_{AB}$ are generators of $so(2,2)$, and $\langle , \rangle$ is a antidiagonal bilinear form on $so(2,2)$ algebra, such that $\langle \Gamma_{AB} ,\Gamma_{CD} \rangle=\epsilon_{ABCD}$, the Newton constant $G$ is taken to be $1$.
 The shell is discretized (represented as an ensemble of $N$ circularly arranged particles)
\begin{equation}\label{actionshell}
S_{shell}=\sum\limits_i^N \int_{l_i} Tr(K_i A_\mu) dx^\mu,
\end{equation}
where $l_i$ is i-th particle worldline and $K_i=m_i \Gamma_{03}$ -- a fixed element of so(2,2)-algebra, corresponding to time translations, $m_i$ is the mass of $i$-th particle.

The symplectic form for particles composing the shell was derived in \cite{aes2}, the derivation closely followed \cite{am},
and the result was:
\begin{equation}
\Omega_i=\langle \delta h_ih_i^{-1},\wedge u_i^{-1}\delta u_i\rangle,
\end{equation}
where $h_i$ is an SO(2,2) transport from the reference point to the i-th particle and $u_i$ is a holonomy around i-th  particle originated
at the same reference point, $u_i=h_i^{-1}\exp(\pi lm_iH)h_i$, where $H=\Gamma_{12}$ is the spatial rotation generator, and $m_i$ is the mass of i-th particle.
In terms of $h_i$ it could be rewritten as
\begin{equation}
\Omega_i=\langle \exp(-\pi l m_iH)  h_i^{-1} \delta h_i \exp(\pi l m_iH),\wedge h_i^{-1}\delta h_i\rangle.
\end{equation}

The symplectic form of the full shell was obtained by composition of holonomies of the constituents, $u=\prod_i u_i$
\begin{equation}\label{shellsf}
\Omega_{shell}=\sum_i \Omega_i =\langle \delta hh^{-1},\wedge u^{-1}\delta u\rangle.
\end{equation}
Here $h=h_0$ and $u=h^{-1}\exp(\pi lMH)h$, $M$ is the total energy of the shell and $H=\Gamma_{12}$ when $M<1$, i.e. the shell collapses to a point particle,
 and $H=\Gamma_{01}$ when $M>1$, i.e. the shell collapses to Ba\~{n}ados-Teitelboim-Zanelli  black hole. In other words $H$ is a Lorentz generator which leaves the singularity worldline stable.

 One can show that $so(2,2)$-algebra is a classical Drinfeld double $D(sl(2))$ \cite{frt}, \cite{dd}, \cite{meus}.
It consists of
Lorentz transformations generated by
$J^a=\epsilon^{abc}\Gamma_{ab}$ and translations
generated  by
$P_a=1/l\Gamma_{a3}$. Commutation relations between them are $[J^a,J^b]=\epsilon^{abc}J_c$,  $[P^a,J^b]=\epsilon^{abc}P_c$
$[P^a,P^b]=\Lambda \epsilon^{abc}J_c$.
One can see that translations do not form a subalgebra.
One can  choose a different basis in which Lorentz transformations and (modified) translations both form subalgebras:
 $$
 x_0=2iJ_1, x_1=-J_0+iJ_2, x_2=J_0+iJ_2.
 $$
 $$
X_0=\frac{1}{2}iP_1, X_1=\frac{1}{2}(P_0+iP_2)+\frac{\Lambda}{2}x_2, X_1=\frac{1}{2}(-P_0+iP_2)-\frac{\Lambda}{2}x_1,
$$
with possible exchange $J_1,P_1 \rightleftarrows J_0,P_0$.
As before, Lorentz transformations form $sl(2)$ subalgebra :
 $$
[x_0,x_1]=2x_1, [x_0,x_2]=-2x_2, [x_1,x_2]=x_0.
$$
Modified translations now also form a subalgebra
$$
[X_0,X_1]=\frac{\Lambda}{2}X_1, [X_0,X_2]=-\frac{\Lambda}{2}X_2, [X_1,X_2]=0,
$$
which is a sum of two Borel subalgebras of $sl(2)$, $B^+\oplus B^- $ with diagonal elements identified.
 Cross commutation relations between new translations and Lorentz transformations
$$
[x^0,X_0]=0, [x^0,X_1]=x^2+\frac{\Lambda}{2}X_1, [x^0,X_2]=-x^1+\frac{\Lambda}{2}X_2,
$$
$$
[x^1,X_0]=2x^1, [x^1,X_1]=-2x^0-\frac{\Lambda}{2}X_0, [x^1,X_2]=0,
$$
$$
 [x^2,X_0]=-2x^2, [x^2,X_1]=0, [x^2,X_2]=2x^0-\frac{\Lambda}{2}X_0.
$$
It leaves Ad-invariant the following bilinear form:
\begin{equation}
\langle x_a,X^b\rangle =\delta_a^b, \ \langle x_a,x^b\rangle =0, \ \langle X_a,X^b\rangle =0.
\end{equation}
This algebra is the classical Drinfeld double $D(sl(2))$.

This can be promoted to a Lie bialgebra with cocommutator given by
\begin{equation}
\delta_D (Y)=[Y\otimes 1 +1 \otimes Y, r], \ \forall Y \in \{ x_a, X_a\},
\end{equation}
where
\begin{equation}\label{crm}
r=\sum_a X_a \otimes x^a
\end{equation}
is the classical r-matrix. It automatiacally satisfies the classical Yang-Baxter equation.
Cocommutator depends on its skew symmetric part,
\begin{equation}
r'=\sum_a X_a \wedge x^a.
\end{equation}
In terms of the initial generators, $J_a$, $P^a$ it can be rewritten as
\begin{equation}
r'= {(\Lambda) J_0\wedge J_2}+
{-P_0 \wedge J_0 +P_1 \wedge J_1 + P_2 \wedge J_2 }.
\end{equation}
Here the first term is the skew  symmetric  part  of   sl(2)  r-matrix and the second term is what survives $\Lambda \rightarrow 0$  limit.

This Lie bialgebra structure can be related to
Poisson-Lie structure on the phase space obtained from symplectic form (\ref{shellsf}).

% On the phase space of the shell a symplectic form has been derived \cite{aes1,aes2}
%\begin{equation}
%\Omega_{shell} = \langle \delta h_0 h_0^{-1},\wedge U^{-1}\delta U\rangle=\langle e^K \delta h_0 h_0^{-1} e^{-K},\wedge \delta h_0 h_0^{-1}\rangle
%\end{equation}
%where $h_0$ is $SO(2,2)$ transformation between a point on the shell and the origin, $K$ is a Lorentz generator which leaves singularity worldline stable, and $U=h_0 %e^K h_0^{-1}$ - the holonomy around the shell.

 Decompose $h=h_L h_T$, where $h_L=\exp(\alpha_a x^a)$ - Lorentz transform and
 $h_T=\exp(\beta_a X^a)$ -modified translation which is a subgroup. Substituting this into (\ref{shellsf}) one obtains:
\begin{equation}\nonumber
\Omega_{shell} = \langle \delta h_T h_T^{-1},\wedge u_L^{-1}\delta u_L\rangle=\langle  \delta h_T h_T^{-1},\wedge h_L^{-1} e^{-K}\delta h_L h_L^{-1} e^K h_L \rangle.
\end{equation}
From this one derives the following Poisson brackets
\begin{equation}
\{ h_T ,\otimes U_L \}= (1\otimes U_L) r (h_T\otimes 1),
\end{equation}
with r-matrix given by (\ref{crm}).
 The infinitesimal version of this Poisson-Lie group is $D(sl(2))$ Lie bialgebra, and its quantization results in quantum double  $D(SL_q(2))$  with $q=\exp(-\pi \sqrt{|\Lambda|} \hbar)$ or $q=\exp(i\pi \sqrt{|\Lambda|} \hbar)$..

\section{Quantum double and *-relations.}\label{sec3}
Quantum double $D(SL_q(2))$ is a unity of quantum universal enveloping algebra,  $U_q(sl(2))$, and its dual, quantized algebra of functions on a group,
$Fun(SL_q(2))$, with cross commutation relations between the two.

 Coordinate space is the algebra of deformed translations in $ADS^3$ space: $U_q(sl(2))$: $X_\pm, q^H$,
\begin{equation}\label{uq}
q^{H/2}X_{\pm}q^{-H/2} = q^{\pm 1}X_\pm, \ \ \ \ [X_+,X_-]=\frac{q^H -q^{-H}}{q-q^{-1}}.
\end{equation}
It contains a central element, quadratic Casimir: $C_2=X^+X^-+\left(\frac{q^{\frac{1}{2}(H-1)}-q^{-\frac{1}{2}(H-1)}}{q-q^{-1}}\right)^2$. Geometrically, it has a meaning of invariant distance in $ADS^3$ and can be identified with the shell radius.

Momentum space is an $SL_q(2)$ holonomy around the shell:
\begin{eqnarray}\nonumber
u=\left(
\begin{array}{c  c}
a & b \\
c & d
\end{array}
\right).
\end{eqnarray}
It is represented by non-commutative algebra $Fun(SL_q(2))$: $a,b,c,d$, $ad-qbc=1$,
\begin{equation}\label{fq}
ab=qba, \ \ \ ac=qca, \ \ \ bd=qdb, \ \ \ cd=qdc,
\end{equation}
\begin{equation}\nonumber
bc=cb, \ \ \ \ \ ad-da=(q-q^{-1}) bc.
\end{equation}
 Cross commutation relations between $U_q(sl(2))$ and $Fun(SL_q(2))$ in the case when $U_q(sl(2))$ acts on $Fun(SL_q(2))$ as left-invariant derivative are
\begin{eqnarray}\nonumber
q^{H^L}a=q^{-1}aq^{H^L}, \ \  q^{H^L}b=qbq^{H^L}, \ \  X_-^La=q^{1/2}aX_-^L, \ \  X_+^Lb=q^{-1/2}bX_+^L, \\ \nonumber
 X_+^La=q^{1/2}aX_+^L+q^{-1/2}bq^{H^L/2}, \ \  X_-^Lb=q^{-1/2}bX_-^L+q^{1/2}aq^{H^L/2},
\end{eqnarray}
plus relations obtained by replacement
 $a\leftrightarrow c$,$b\leftrightarrow d$.

When $U_q(sl(2))$ acts on $Fun(SL_q(2))$ as right-invariant derivative the cross commutation relations are
\begin{eqnarray}\nonumber
q^{H^R}a=q^{-1}aq^{H^R}, \ \  q^{H^R}b=q^{-1}cq^{H^R}, \ \  X_-^Ra=q^{1/2}aX_-^R, \ \  X_+^Rc=q^{-1/2}cX_+^R \\ \nonumber
 X_+^Ra=q^{1/2}aX_+^R+q^{-1/2}cq^{H^R/2}, \ \  X_-^Rc=q^{-1/2}cX_-^R+q^{1/2}aq^{H^R/2},
\end{eqnarray}
plus relations obtained by replacement
 $a\leftrightarrow b$,$c\leftrightarrow d$.

The above commutation relations hold both for $SL(2)$ and for $SU(1,1)$ case.
The difference between the two is in *-relations (real forms).

 Depending on the total energy the shell collapses either to point particle with trajectory along $P_0$, or to BTZ black hole with trajectory
along $P_1$.
In the case of point particle the singularity is timelike.
The symplectic form looks like:
\begin{equation}
\Omega_{shell} =\langle e^{iJ_0} \delta h h^{-1} e^{i\pi \sqrt{|\Lambda|}J_0},\wedge \delta h h^{-1}\rangle.
\end{equation}
the role of $H$ in (\ref{uq}) is played by $iP_0$. This is $D(SU_q(1,1))$-case with the following *-relations:
$D(SU_q(1,1))$-case, $H=iP_0$, $q$-real
\begin{equation}\nonumber
a^*=d, \ \ b^*=qc, \ \ H^*=H, \ \ X_\pm^*=X_\mp, \ \ q^*=q.
\end{equation}
Here $q$ is real.

In the case of BTZ black hole the singularity is spacelike.
The symplectic form looks like:
\begin{equation}
\Omega_{shell} =\langle e^{iJ_1} \delta h h^{-1} e^{i\pi \sqrt{|\Lambda|}J_1},\wedge \delta h h^{-1}\rangle.
\end{equation}
he role of $H$ in (\ref{uq}) is played by $iP_1$. This is $D(SL_q(2))$-case with the following *-relations:
\begin{equation}\nonumber
a^*=a, \ \ b^*=b, \ \ H^*=-H, \ \ X_\pm^*=-X_\pm, \ \ q^*=q^{-1}.
\end{equation}
Here $q$ is a root of unity.

\section{Representations and spectra}\label{sec4}
Following the previous work \cite{aes1,aes2} we shall try to construct
momentum representation of the above algebra.

The momentum variables are now non-commutative.
So the states are no longer wave functions, but ordered polynomials acting on $\mathbf{1}$:
\begin{equation}
\Psi = \sum \alpha_{klmn} a^k b^l c^m d^n \mathbf{1}.
\end{equation}
If $k,l,m,n \geq 0$ these states are not normalizible (as it could be seen in $q\rightarrow 1$ limit).
These states correspond to finite dimensional non-unitary representations of $sl(2)$.
In our case, the states of interest are those which in the limit $q\rightarrow 1$ correspond to infinite-dimensional unitary representations \cite{gn }. These states are
normalizible and need to contain negative degrees of combinations of $a,b,c,d$ which are invertible.

 In $SU_q(1,1)$ case $aa*=ad=1+bb*\geq 1$, so $a$ is invertible and the same is true  for $d$.
 The lowest weight states (those which are annihilated by $X_+$ or $X_-$) are
\begin{equation}
\Psi_{n,n} = a^{-n} \mathbf{1}, \ \ \ \Psi_{n,-n} = d^{-n} \mathbf{1}.
\end{equation}
The rest of states is obtained by applying $X_{\pm}$ to them. The time operator $T=H$ has discrete, but unbounded spectrum, $T\Psi_{n,n}=n\Psi_{n,n}$.
The structure of representations is the same as in $q=1$ limit. But this is  a  no black hole case.

Black hole case corresponds to $SL_q(2)$, which for $q \neq 1$ is not isomorphic to $SU_q(1,1)$ (see \cite{frt}), but in the limit $q=1$ the isomorphism has to be restored.

Introduce new variables which resembles $SL(2) \rightarrow \ SU(1,1)$ transformation
$$
\tilde a =q^{-1/2}a+ib-ic+q^{1/2}d,
$$
$$
\tilde b =-q^{-1/2}a+ib+ic+q^{1/2}d,
$$
$$
\tilde c = q\tilde b^*,
$$
$$
\tilde d = \tilde a^*.
$$
Of cause, commutation relations of the type (\ref{fq}) will not hold for $\tilde a,\tilde b,\tilde c,\tilde d$.

Notice that $\tilde a$ is an invertible combination  of $SL_q(2)$ elements, because
$$
\tilde a^* \tilde a = q+q^{-1}+a^2+b^2+c^2 +d^2>1.
$$
The same is true for $\tilde d$.
 Combinations of the type
$$
q^{(-\alpha^L H^L-\alpha^R H^R)}\tilde a q^{(\alpha^L H^L+\alpha^R H^R)} =
$$
$$
=q^{1/2+\alpha^L+\alpha^R}a-iq^{-\alpha^L+\alpha^R}b+iq^{\alpha^L-\alpha^R}c+q^{-1/2-\alpha^L-\alpha^R}d,
$$
where $\alpha^L$ and $\alpha^R$ are arbitrary real numbers, are also invertible because $q^{(-\alpha^L H^L-\alpha^R H^R)}$ is invertible.

Introduce
$$
\tilde H^L =i(q^{-1/2}X_+^Lq^{-H^L/2} - q^{+1/2}X_-^Lq^{-H^L/2})
$$
and
$$
\tilde H^R =i(q^{-1/2}X_+^Rq^{-H^R/2} - q^{+1/2}X_-^Rq^{-H^R/2}).
$$
These resembles $su(1,1)$ Cartan operator in $q\rightarrow 1$ limit.  The specific form chosen is motivated by the fact that
 $\tilde a^{-1} \mathbf{1}$ appears to be an eigenstate of both $\tilde H^L$ and $\tilde H^R$ with the same eigenvalue, $\tilde H^L\tilde a^{-1} \mathbf{1}=\tilde a^{-1} \mathbf{1}$, $\tilde H^R\tilde a^{-1} \mathbf{1}=\tilde a^{-1} \mathbf{1}$.
By $q\rightarrow 1$-correspondence one can identify them with time operator $T=\tilde H^L$, and angular momentum operator $J= \tilde H^L-\tilde H^R$, (see for example \cite{mw}). They are both hermitian $T^*=T$, $J^*=J$. Cylindrically symmetric states correspond to $J=0$, i.e.  $\tilde H^L=\tilde H^R$.

The eigenstates of time operator  $T=\tilde H^L$ can be obtained by recursion formula:
$$
 \Psi_{n+1}= (\alpha_n(q^{-1/2}a-ic)+ib+q^{1/2}d) \Psi_{n},
$$
where $\alpha_n$ are  unknown coefficients to be found. Applying $\tilde H^L$ to it leads to the following equations:
$$
\lambda_{n+1}-q\lambda_n=\alpha_n^{-1}, \ \ \lambda_{n+1}-q^{-1}\lambda_n=\alpha_n.
$$
They have two solutions $\alpha_n =q^n$ and $\alpha_n =-q^{-n}$ and the eigenvalues are found to be
$\lambda_n=\frac{q^{ n}-q^{- n}}{q-q^{-1}}$. Of the two solutions the first by $q\rightarrow 1$ correspondence can be identified with lowest weight states.
To obtain cylindrically symmetric states they have to be symmetrized with respect to left and right action of $U_q(sl(2))$.
As a result, the lowest weight cylindrically symmetric  normalizible states are:
\begin{equation}\nonumber
\Psi_{l,l}  =  \prod_{k=1}^{l}q^{-\frac{k}{2}( H^L+ H^R)}\tilde a q^{\frac{k}{2}( H^L+ H^R)} {\bf 1} =\prod_{k=1}^{l} \Big(q^k(q^{-1/2+k}a-ic)+(iq^kb+q^{1/2}d) \Big)^{-1} {\bf 1},
\end{equation}
\begin{equation}\nonumber
\Psi_{l,-l}  =  \prod_{k=1}^{l}q^{-\frac{k}{2}( H^L+ H^R)}\tilde d q^{\frac{k}{2}( H^L+ H^R)} {\bf 1}  = \prod_{k=1}^{l} \Big(q^{k}(q^{1/2+k}a+ic)+(-iq^kb+q^{-1/2}d) \Big)^{-1} {\bf 1}.
\end{equation}
 The eigenvalues of time operator
\begin{equation}\nonumber
T \Psi_{l,\pm l}  =\frac{q^{\pm l}-q^{\mp l}}{q-q^{-1}} \Psi_{l,\pm l}=[\pm l]_q \Psi_{l,\pm l}.
\end{equation}
 Unlike $SU_q(1,1)$, the eigenvalues of time operator are now q-integers.
 One can show that these states are also eigenstates  of the quadratic Casimir operator:
 \begin{equation}\nonumber
  C_2\Psi_{l,n}=((q^{\frac{1}{2}(l-1)}-q^{-\frac{1}{2}(l-1)})/(q-q^{-1}))^2 \Psi_{l,n}.
\end{equation}

To obtain the rest of states it would be convenient to find rising and lowering operators. Because the spectrum of the Cartan operator (time operator in our case) is
not equidistant, commutation relations are sought for in most general form:
\begin{equation}\nonumber
T \tilde X_{\pm,n} - X'_{\pm,n} T = \lambda_{n\pm 2} X_{\pm,n} -  \lambda_{n} X'_{\pm,n},
\end{equation}
where $X_\pm$ is rising (lowering) operator and   $X'_\pm$ some other operator. One can show that this equation has solution only if $\lambda_n=\frac{q^{ n}-q^{- n}}{q-q^{-1}}$, as before, and the solution is
\begin{equation}\nonumber
 \tilde X_{\pm,n} = i(q^{\mp n +1/2} X_+ + q^{\pm n -1/2}X_-)q^{H/2}\pm \frac{q+q^{-1}}{q-q^{-1}}(q^H-1),
\end{equation}
\begin{equation}\nonumber
 \tilde X'_{\pm,n} = i(q^{\mp n +1/2\mp2} X_+ + q^{\pm n -1/2\pm2}X_-)q^{H/2}\pm \frac{q+q^{-1}}{q-q^{-1}}(q^H-1)=\tilde X_{\pm,n\pm 2}.
\end{equation}
These operators were derived for left invariant action of $U_q(sl(2))$ on $Fun_q(SL(2))$. For right invariant action the equations are fully analogous.
Notice that operators $\tilde X_{\pm,n}$ depend on the level $n$, there is no universal rising and lowering operators for every level.

The remaining states can be constructed using the above operators. For cylindrical symmetry they have to be symmetrized with respect to left invariant and right invariant action:
\begin{equation}\nonumber
\Psi_{l,l+2n}  = \prod_{k=1}^{n}( \tilde X^L_{+,l+2(k-1)} \tilde X^R_{+,l+2(k-1)} )\Psi_{l,l},
\end{equation}
and
\begin{equation}\nonumber
\Psi_{l,-l-2n}  = \prod_{k=1}^{n}( \tilde X^L_{-,l-2(k-1)} \tilde X^R_{,l-2(k-1)} )\Psi_{l,-l}.
\end{equation}
Obviously they remain eigenstates of the Casimir operator with the same eigenvalue,
\begin{equation}\nonumber
  C_2\Psi_{l,n}=((q^{\frac{1}{2}(l-1)}-q^{-\frac{1}{2}(l-1)})/(q-q^{-1}))^2 \Psi_{l,n}.
\end{equation}
The eigenvalues of time operator are then
\begin{equation}\nonumber
T \psi_{l,n}=[n]_q\Psi_{l,n}.
\end{equation}
Because $q$ is a root of unity the eigenvalues of time operator and the Casimir operator change periodically with $n$.
Half of a period of the eigenvalues of the Casimir operator corresponds to the full period of the eigenvalues of the time operator.  Negative eigenvalues of the time operator correspond to the white hole part of the Penrose diagram.

Notice that unlike $SU_q(1,1)$ case with real deformation parameter, the eigenvalues of time operator are now quantum integers. Given that $[N]_q=0$,  where $N=1/(\sqrt{|\Lambda|} \hbar)$the representation becomes finite-dimensional, with dimension equal $N-l$. The number of representations themselves is also finite $N$.

\begin{equation}
l=0...N, \ \ \ n=-N..-l,l..N, \ \ \ N=1/(\sqrt{|\Lambda|} \hbar)\ \ \ q^N=-1
\end{equation}
Inside the black hole $n$ and $l$ vary within a finite range, so the Hilbert space is finite-dimensional.

\section{Discussion}

When the spectrum of states inside the black hole is found, one can ask about transition amplitudes between these states. This will require calculation of matrix elements of the evolution operator. This, in turn, needs the development of integral calculus on quantum groups.

Although, to our knowledge, the integral calculus on non-compact quantum groups at a root of unity in general form is lacking, there may be recursion relation between matrix elements of interest. This work is now in progress.

Already at this stage one can argue that all the above matrix elements will be finite. This will follow from the boundeness of the evolution operator and normalizibility of the states found.

As to what happens to the shell at the central singularity: the subspace of the Hilbert space corresponding to $R=0$ is one dimensional, there is only one possible eigenvalues of time operator. This means that the shell cannot rest at the singularity and in the next moment of time will be out of there. This can be interpreted as a quantum bounce at the singularity. The shell will enter the white hole region of the Penrose diagram, and then cross the horizon in the opposite direction.

Given the main result of this paper, that the Hilbert space of a black hole is finite dimensional, one can ask if its dimensionality could be related to the black hole entropy. The answer is no: the spherically symmetric states considered here are not enough to form the black hole entropy, their number is exponentially small compared to what is needed for recovering the Bekenstein-Hawking formula. For calculating entropy one has to give up spherical symmetry.

\bigskip

{\bf Acknowledgements.}
We are grateful to V. Berezin for explanation of his work and to Y. Elmahallawy for collaboration on the earlier papers. A.A. was supported by  RSF Grant No. 21-12-00020, A.S. is grateful to L.Euler international mathematical institute (grant 075-15-2022-289) for support of this work.

\end{document}